\documentclass[a4paper]{article}
\usepackage{INTERSPEECH2020}
\usepackage{amsmath,graphicx}
\usepackage{multirow}
\usepackage{graphicx}
\usepackage{hyperref}
\usepackage{color}
\usepackage{subcaption}
\newcommand{\bs}[1]{\boldsymbol{#1}}
\title{Improved Prosody from Learned F0 Codebook Representations for VQ-VAE Speech Waveform Reconstruction}

\name{Yi Zhao$^1$,  Haoyu Li$^1$,  Cheng-I Lai$^2$, Jennifer Williams$^{3}$, Erica Cooper$^1$, Junichi Yamagishi$^{1,3}$}

\address{
  $^1$National Institute of Informatics, Japan
  $^2$Massachusetts Institute of Technology, USA \\
 $^3$University of Edinburgh, UK
}
\email{$\{$zhaoyi,haoyuli,ecooper,jyamagis$\}$@nii.ac.jp, clai24@mit.edu, j.williams@ed.ac.uk}

\begin{document}
%
\maketitle
\begin{abstract} 
Vector Quantized Variational AutoEncoders (VQ-VAE) are a powerful representation learning framework that can discover discrete groups of features from a speech signal without supervision. Until now, the VQ-VAE architecture has previously modeled individual types of speech features, such as only phones or only F0. This paper introduces an important extension to VQ-VAE for learning F0-related suprasegmental information simultaneously along with traditional phone features. The proposed framework uses two encoders such that the F0 trajectory and speech waveform are both input to the system, therefore two separate codebooks are learned. We used a WaveRNN vocoder as the decoder component of VQ-VAE. Our speaker-independent VQ-VAE was trained with raw speech waveforms from multi-speaker Japanese speech databases. Experimental results show that the proposed extension reduces F0 distortion of reconstructed speech for all unseen test speakers, and results in significantly higher preference scores from a listening test. We additionally conducted experiments using single-speaker Mandarin speech to demonstrate advantages of our architecture in another language which relies heavily on F0. 
%
\end{abstract}
\noindent\textbf{Index Terms}: VQ-VAE, speech synthesis, prosody, representation learning
\vspace{-1.5mm}
\section{Introduction}
\vspace{-1mm}
\label{sec:intro} 

Speech signals contain rich factors existing at different linguistic layers such as prosody, content, and timbre simultaneously. Modelling and controlling these factors through neural representation learning is essential for many speech applications~\cite{Chung2018,oord2018representation,chorowski2019unsupervised}. 
In this paper, we focus on Vector Quantized Variational AutoEncoder (VQ-VAE)~\cite{van2017neural} since it is a promising self-supervised representation learning approach suitable for tasks such as voice conversion (VC)~\cite{van2017neural,Ding2019}, text-to-speech (TTS) synthesis~\cite{henter2018deep,wang2019vector,tjandra2019VQVAE}, speech coding~\cite{garbacea2019low}, and even music generation 
\cite{dhariwal2020jukebox} wherein we need to model and control content, prosody and speaker characteristics separately. 

The VQ-VAE paradigm typically consists of three main components: an encoder network, VQ codebooks, and a decoder network. For speech-related applications, the expected role of the encoder is to extract phone- or syllable-equivalent segmental representations, while the decoder reconstructs the input raw speech waveform via a neural vocoder, such as WaveNet~\cite{oord2016wavenet} or WaveRNN~\cite{kalchbrenner2018efficient}. Between the encoder and decoder is a trainable VQ codebook. The VQ codebook transforms outputs of the encoder network into a set of discrete representations for the down-stream task. 

A well-known application of VQ-VAE is voice conversion, which is done by conditioning the decoder with a speaker one-hot vector or a speaker-embedding vector during training, and then simply swapping the speaker vector to convert the speaker vector to a different speaker during inference. In~\cite{Ding2019}, promising conversion results using an English speech database have been reported. However, suprasegmental features such as F0, another important cue in speech signals, are not properly modelled in the voice conversion framework. 

From a preliminary experiment, we found that VQ-VAE-based speech waveforms typically have inappropriate prosodic structure in the case of Japanese. This is probably because Japanese is a pitch-accented language, which means that pitch accents directly affect the meaning of words and the perceived naturalness of the speech. However, unlike tonal languages such as Mandarin, in which each syllable is coupled with a specific tone index, Japanese pitch accentual patterns (high or low) of each mora are affected by the information at a different linguistic layer from the syllable layer, such as adjacent words. Hence, we hypothesize that a successful VQ-VAE architecture needs to simultaneously extract not only representations corresponding to the segmental features, but also another set of representations corresponding to the supra-segemental features. 

Motivated by this, we propose an extension to VQ-VAE structure utilizing two encoders at the same time. One encoder uses a raw speech waveform for input exactly as the original VQ-VAE does. The other encoder uses F0 trajectory as input to separately learn the pitch patterns as well as other F0-related supra-segmental information. This model was trained using a loss function that jointly considers two types of VQ losses as well as the usual coarse-to-fine waveform losses used for training WaveRNN. Listening test results show that this simple yet effective extension significantly improves prosody and naturalness of reconstructed Japanese speech waveforms. The extended VQ-VAE structure was motivated by Japanese prosody, but it can be applied to speech in any language. Therefore, we also show results of the extended VQ-VAE using a Mandarin speech database for further analysis.   

This paper is structured as follows: Section 2 gives an overview of the original VQ-VAE framework, and Section 3 explains our proposed extension for the F0 representation. Section 4 elaborates the details of the extended VQ-VAE. Section 5 shows experimental results in Japanese and Mandarin. Finally, we summarize our findings in Section 6. 

\vspace{-1.5mm}
\section{Overview of VQ-VAE}
\label{sec:VQ-VAE} 
\vspace{-1mm}

VQ-VAE is a self-supervised learning technique that uses an encoder network, VQ codebooks, and a decoder network, as defined below: 
%
\begin{align}
\vspace{-2mm}
     \bs{z}_{1:N} &= \text{Encoder}_{\bs{\Phi}_1}(\bs{o}_{1:T}),  \\
     \bs{e}_{1:N} &= \text{Vector\_quantization}_{\bs{\Phi}_2}({\bs{z}_{1:N}}), \\
     \bs{\widehat{o}}_{1:T} &= \text{Decoder}_{\bs{\Phi}_3}(\bs{e}_{1:N}, \bs{s})
\vspace{-2mm}
\end{align}
The encoder $\bs{\Phi}_1$ takes a raw speech waveform of length $T$, $\bs{o}_{1:T}=\{\bs{o}_1, \cdots, \bs{o}_T\}$, as the input and first encodes it into a raw latent vector sequence $\bs{z}_{1:N}=\{\bs{z}_1, \cdots, \bs{z}_N\}$. Using the raw latent vectors, the vector quantization function $\bs{\Phi}_2$ returns a quantized code vector (that is, centroid) sequence $\bs{e}_{1:N}=\{\bs{e}_1, \cdots, \bs{e}_N\}$. Down-stream tasks can use \textit{indices} of the quantized code vectors as categorical representations. For example, in TTS the indices would represent pseudo phone symbols. Finally, the decoder network $\bs{\Phi}_3$ such as WaveNet or WaveRNN reconstructs the speech waveform of the same length $T$ using the quantized code sequence $\bs{e}_{1:N}$. The decoder network can optionally include a speaker vector $\bs{s}$ as a global condition.  

This VQ-VAE is trained using a penalized log-likelihood function below: 
\begin{equation}
\begin{split}
\vspace{-2mm}
\mathcal{L}&(\bs{\Phi})= -\log{p(\bs{o}_{1:T} | \bs{e}_{1:N}; \bs{\Phi}_3)} \\
&+{\big \Vert}\bs{e}_{1:N} - \text{sg}[\bs{z}_{1:N}]{\big \Vert}^2_2 + \beta{\big \Vert}\bs{z}_{1:N} - \text{sg}[\bs{e}_{1:N}]{\big \Vert}^2_2, 
\label{eq:VQ-VAE_VQ-VAE}
\end{split}
\vspace{-3mm}
\end{equation}
The first term is a log-likelihood function of the decoder network that measures appropriateness of reconstructed speech waveforms. In the case of WaveRNN, this term corresponds to the multi-class cross-entropy loss using a dual softmax layer that predicts the coarse- and fine-waveform representations~\cite{kalchbrenner2018efficient}. The second term drives the quantized vectors towards the raw latent vectors. The third term is a penalty term that prevents the output of the encoder from growing arbitrarily large~\cite{van2017neural}. The operator $\text{sg}[\cdot]$ zeroes out the gradient back-propagated to the argument. $\beta$ is a hyper-parameter corresponding to the commitment loss to make sure the encoder commits to the codebook embedding.

\vspace{-1.5mm}
\section{Proposed Extension for F0}
\vspace{-1mm}

\begin{figure}[t]
\centering
\includegraphics[width=0.9\columnwidth]{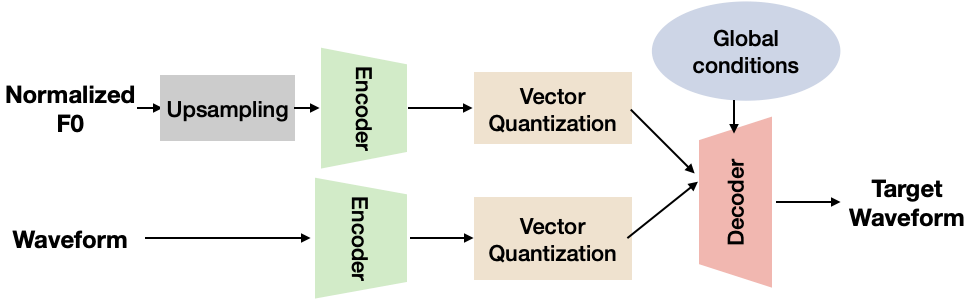}
\vspace{-2mm}
\caption{Overall framework}\label{fig:framework}
\vspace{-5mm}
\end{figure}

In this paper, we extend the VQ-VAE structure and introduce two encoders. One of them uses a raw speech waveform $\bs{o}_{1:T}$ of length $T$ as the input, as the original VQ-VAE does. The other encoder uses an upsampled F0 trajectory $\bs{F}_{1:T}=\{F_1, \cdots, F_T\}$ of length $T$ as the input, as shown in Figure \ref{fig:framework}:
\begin{align}
\vspace{-1.5mm}
     \bs{z}^{o}_{1:N} &= \text{Encoder}_{\bs{\Phi}_{1_o}}(\bs{o}_{1:T}), \label{eqn:wave-enc}\\
     \bs{z}^{F}_{1:N} &= \text{Encoder}_{\bs{\Phi}_{1_F}}(\bs{F}_{1:T}), \label{eqn:f0-enc} \\
     \bs{e}^{o}_{1:N} &= \text{Vector\_quantization}_{\bs{\Phi}_{2_o}}({\bs{z}^{o}_{1:N}}), \label{eqn:wave-vq}\\
     \bs{e}^{F}_{1:N} &= \text{Vector\_quantization}_{\bs{\Phi}_{2_F}}({\bs{z}^{F}_{1:N}}), \label{eqn:f0-vq} \\
     \bs{\widehat{o}}_{1:T} &= \text{Decoder}_{\bs{\Phi}_3}(\bs{e}^{o}_{1:N},\, \bs{e}^{F}_{1:N}, \bs{s}) \label{eqn-decoder}
\vspace{-2mm}
\end{align}
Outputs of the two encoders are separately quantized and the speech waveform is then reconstructed using the two sets of code vectors. In the extended VQ-VAE, two sets of representations, i.e., indices of the quantized code vectors for each encoder become available for the down stream tasks. We hope that the discrete representations learned from F0 capture pitch patterns and/or other suprasegmental information. The proposed extension is straightforward, but we observe that it results in impressive improvements for prosody, especially F0 of reconstructed speech waveforms in Japanese. 

Its loss function can also be defined in a similar way:  
\begin{equation}
\vspace{-1.5mm}
\begin{split}
\mathcal{L}&(\bs{\Phi})= -\log{p(\bs{o}_{1:T} | \bs{e}^{o}_{1:N}, \bs{e}^{F}_{1:N}; \bs{\Phi}_3)} \\
&+{\big \Vert}\bs{e}^o_{1:N} - \text{sg}[\bs{z}^o_{1:N}]{\big \Vert}^2_2 + \beta{\big \Vert}\bs{z}^o_{1:N} - \text{sg}[\bs{e}^o_{1:N}]{\big \Vert}^2_2, \\
&+\gamma({\big \Vert}\bs{e}^F_{1:N} - \text{sg}[\bs{z}^F_{1:N}]{\big \Vert}^2_2 + \beta{\big \Vert}\bs{z}^F_{1:N} - \text{sg}[\bs{e}^F_{1:N}]{\big \Vert}^2_2), 
\label{eq:loss}
\end{split}
\vspace{-2mm}
\end{equation}
Here $\gamma$ is a new hyper-parameter. 

This extension is related to VQ-VAE based speech coding by G{\^a}rbacea \MakeLowercase{\textit{et al.}}\ \cite{garbacea2019low} wherein a second \textit{decoder} was introduced to predict both a speech waveform and an F0 trajectory at the same time. F0 prediction loss was also introduced in the training process in \cite{garbacea2019low}. However, unlike our extension, a common encoder and codebooks were used.
Another related extension comes from \textit{SPEECHSPLIT} \cite{qian2020unsupervised} wherein three different auto-encoders were used to disentangle the speech signal into components, however their aim was to extract continuous latents. 

\vspace{-1.5mm}
\section{Details of the Extended VQ-VAE}
\vspace{-1mm}

We implemented the extended VQ-VAE based on a public implementation of VQ-VAE\footnote{\url{https://github.com/mkotha/WaveRNN}}. Here we describe details of the extended VQ-VAE used for our experiments. 




\vspace{-1.5mm}
\subsection{Waveform and Upsampled Normalized F0}
\vspace{-1mm}
 The input waveform representation is linear PCM. 
The output waveform is parameterized for coarse and fine parts separately unlike the input waveform, and they are notated as $\bs{\widehat{o}_t}=(c_t, f_t)$ in the following sections. Linear F0 including unvoiced regions is extracted frame-by-frame and then normalized sentence by sentence using the minimum and maximum values of each utterance.
This normalization helps the model to focus on the pitch trajectory while casting away speaker information. It can also benefit the discretized latent representation learning since it shrinks the dynamic range of continuous F0 values.
Then, the extracted F0 is upsampled via a transposed convolutional layer in order to be aligned with the waveform points. In our implementation, the input waveform is 16-bit linear PCM at 22.05k sampling frequency, and $c_t$ and $f_t$ are both 8-bit.
 
\vspace{-1.5mm}
\subsection{Encoder}
\vspace{-1mm}
\begin{figure}[t]
  \begin{subfigure}{0.48\columnwidth}
     \centering
    \includegraphics[clip, width=0.85\linewidth]{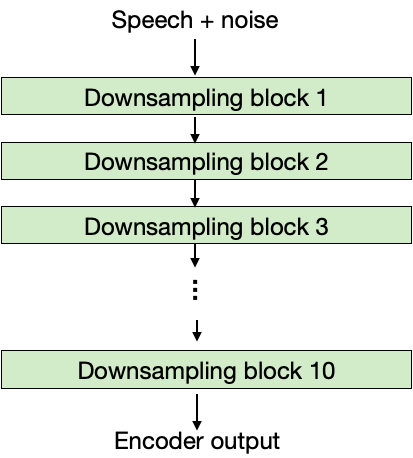}
     \caption{}
  \end{subfigure} \hfill
  \begin{subfigure}{0.48\columnwidth}
     \centering
     \includegraphics[clip,width=0.66\linewidth]{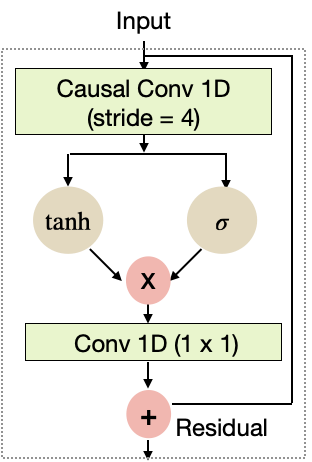}
     \caption{}
  \end{subfigure}
  \vspace{-2mm}
  \caption{Encoder framework (a) and downsampling block (b)}
  \label{fig:encoder}
  \vspace{-5mm}
\end{figure}

Encoders for raw waveform and F0 (Eqs.\ \ref{eqn:wave-enc} and \ref{eqn:f0-enc}) share the same architecture. Each of them has ten down-sampling blocks (Figure \ref{fig:encoder} (a)). Each block consists of a 1D convolution layer, followed by another 1D convolution layer and a gated activation layer using tanh and sigmoid functions. Outputs of the gated activation are further filtered using 1D convolution. The number of channels is set to 256 for the first convolution layer, and 128 for the second convolution layer. The final 1D convolution is also set to 128 channels. The residual connection combines sparsified block input with block output to retain block input information before down sampling. Its diagram is shown in Figure \ref{fig:encoder} (b). To encourage generalization, we introduce additive and multiplicative Gaussian noise to the waveform and F0 inputs. Without these types of noise, the model tends to be sensitive to small input perturbations, and over-fitting occurs after a certain point in training.
The downsampling rate of the entire encoder is approximately $N/T\approx 0.0145$. 




\vspace{-1.5mm}
\subsection{Vector Quantization}
\vspace{-1.5mm}

In vector quantization of Eq.\ \ref{eqn:wave-vq}, the encoder output $\bs{z}^{o}_{n}$ at time step $n$ is quantized using the closest vector included in the VQ codebook, as computed by Euclidean distance.
The closest code vector $\bs{e}^{o}_{n}$ for each time step $n$ is repeated for the entire sequence to obtain $\bs{e}^{o}_{1:N}$. The same operation is conducted for Eq.~\ref{eqn:f0-vq}, with separate codebook for F0 to output $\bs{e}^{F}_{1:N}$. In our implementation, the waveform codebook consists of 512 code vectors of 128 dimensions. As for the F0 codebook, we varied the number of code vectors from 6 to 512 and found out that ten code vectors are capable enough to capture the F0 trajectory variation. The dimension of the F0 code vector is also 128.


\vspace{-1.5mm}
\subsection{Decoder}
\vspace{-1.5mm}

\begin{figure}[t]
\centering
\includegraphics[width=1\columnwidth]{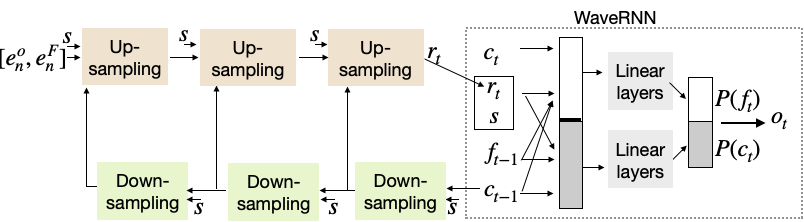}
\caption{Decoder framework. $\bs{r}_t$ is output of the last upsampling block. $c_t$ and $f_t$ represent coarse and fine bits, respectively, of an output waveform $\bs{o}_t$ at time $t$. $P(c_t)$ and $P(f_t)$ at the outputs are probabilities estimated by the dual softmax layer. $\bs{s}$ is the global condition.}\label{fig:decoder}
\vspace{-5mm}
\end{figure}

The purpose of the decoder is to reconstruct a speech waveform $\bs{\widehat{o}}_{1:T}$ using the code vector sequences $\bs{e}^{o}_{1:N}$ and $\bs{e}^{F}_{1:N}$ together with the global condition $\bs{s}$. As shown in Figure~\ref{fig:decoder}, our decoder of Eq.\ \ref{eqn-decoder} was implemented using a combination of upsampling blocks, downsampling blocks, and a WaveRNN module. 

\vspace{0.1mm}
\noindent 
\textbf{Up-sampling block:} Time scales for the code vectors and waveform are different from each other because the encoder downsamples the code vectors. Therefore we use upsampling blocks so that code vectors ($\bs{e}^{o}_{n}$ and $\bs{e}^{F}_{n}$) are at the same operating rate as the waveform points. Each upsampling block is a GRU layer followed by a transposed convolution network. 

\noindent 
\textbf{Down-sampling block:}
Since WaveRNN is an auto-regressive (AR) model, it uses the coarse component $c_{t-1}$ of the output waveform $\widehat{\bs{o}}_{t-1}$ at time $t-1$ as an additional condition for predicting a next waveform point $\widehat{\bs{o}}_{t}$ at time $t$ \cite{kalchbrenner2018efficient}. To do this, the past coarse components are downsampled and combined with the next code vectors. The downsampling block is similar to Figure~\ref{fig:encoder} (b) except for the lack of residual connections. There are three downsampling blocks. The output of each one is connected to the corresponding upsampling block like a U-net \cite{ronneberger2015u}. For the AR feedback, teacher-forcing strategy is used for training and predicted waveforms are used for inference. 

\vspace{0.1mm}
\noindent 
\textbf{WaveRNN module:}
This module takes the output of the last upsampling block $\bs{r}_t$ and the previous sample $\widehat{\bs{o}}_{t-1}=(c_{t-1}, f_{t-1})$, and predicts coarse and fine bits at time $t$ via separated softmax layers, not simultaneously but one-by-one. It first predicts coarse bits $c_t$, then uses $c_t$ to predict fine bits $f_t$. It consists of one common GRU layer and two separated feedforward layers for coarse and fine bits, respectively. In our framework, the global condition $\bs{s}$ is inserted as an additional input to all layers in the decoder.

\vspace{-1.5mm}
\subsection{Global Conditions and Neural Speaker Embedding}
\vspace{-1.5mm}

We used three types of global conditions, gender, emotion, and speaker, as $\bs{s}$ of Eq.\ \ref{eqn-decoder}. The gender and emotions are represented using simple one-hot vectors (two and four dimensions, respectively). It is ideal to determine number of emotional classes from data in an unsupervised way \cite{henter2018deep}, but, for simplicity, we used labeled data in this work. The speaker condition is represented using a neural embedding vector obtained from a speaker encoder, which was pre-trained using the Learnable Dictionary Encoding (LDE) and angular softmax \cite{cai2018exploring}. The original dimension of the speaker embedding vector is 512 and we reduced its dimension to 50 using Linear Discriminant Analysis.\footnote{To consider the balance of gender, emotion, and speaker, we duplicated the gender embedding by 5 times and the emotion embedding by 10 times. Thus the total dimension of global conditions is 100.}. For more details, refer to Section 2 of \cite{cooper2020zeroshot} and its implementation\footnote{\url{https://github.com/jefflai108/pytorch-kaldi-neural-speaker-embeddings}}.

\vspace{-1.5mm}
\subsection{Expected Down-Stream Tasks}
\vspace{-1.5mm}

An example of expected down-stream tasks using our extended VQ-VAE is TTS without any manual labels. Since the proposed model allows us to automatically learn segmental and F0-related suprasegemental discrete representations, it is expected that we can build a TTS system without using manually-defined phone sets or pitch accent labels by just predicting the code indices from text. This would be useful for under-resourced dialects and languages. For instance, Japanese dialects have different pitch accent patterns from the standard ones and there are no established methods for annotating them. Some African languages such as Ibibio also use tones \cite{EKPENYONG2014243}. However, the construction and evaluation of down-stream tasks are beyond the scope of this paper. The focus of this paper is to show that the proposed extended VQ-VAE can produce a higher quality of speech thanks to the F0 encoder and its codebook. 

\vspace{-1.5mm}
\section{Experiments}
\label{sec:exp_con}
\vspace{-1.5mm}

\noindent 
\textbf{Speech databases:} We mainly experimented with Japanese speech databases. We combined the JVS corpus~\cite{takamichi2019jvs}, JTES corpus~\cite{kosaka2018acoustic}, and an in-house corpus introduced in our previous work~\cite{Zhao2019} in order to augment training data. Details of the corpora are as follows: There are a total of 212 speakers (half male and half female), each with around 100 utterances. We used 192 speakers out of the 212 speakers for training and validation. The total number of utterances in the training set is 20,000. The JTES corpus and in-house corpus contain four kinds of emotions: neutral, happy, joy, and angry. The JVS corpus has only neutral speech. 
We selected 8 speakers and 10 utterances per speaker as the test set. Our test speakers are completely unseen and omitted from both training and validation sets. The testing utterances were common across all speakers.
In addition, we conducted a small experiment on a Mandarin dataset to analyze the performance in another tonal language. We used a publicly-available single-speaker Chinese standard Mandarin speech corpus\footnote{\url{https://www.data-baker.com/open_source.html}}. Around 9,000 utterances were used for training and validation. 80 utterances were used for evaluation.

All speech data was down-sampled to 22.05kHz and precision was converted to 16 bits per sample. The waveform amplitude was normalized to -26 dBov in advance using ITU-T G.191 called ``sv56''~\cite{sv56}. The silence segments were also trimmed in advance using Librosa~\cite{mcfee2015librosa}.
F0 values were automatically extracted using a CREPE model \cite{kim2018crepe} with a frame shift of 5ms.

\noindent 
\textbf{Speaker encoder and speaker embedding:} Before training VQ-VAE models, we first trained an LDE-based speaker encoder 
using a multi-speaker Japanese corpus called ATR-APP\footnote{\url{https://www.atr-p.com/products/sdb.html}} to construct a language-dependent speaker encoder. 
This corpus has thousands of Japanese speakers and we used 135k utterances for training. Using this pre-trained speaker encoder, we extracted speaker embedding vectors from the training set of the above Japanese databases.




\noindent 
\textbf{VQ-VAE training:} 
Both the original and extended VQ-VAE parameters are optimized using the Adam~\cite{kingma2014adam}.
Hyper-parameters $\beta$ in Eq.~\ref{eq:loss} is initialized as 0.001 and linearly increased to 0.01 after 1k steps.
$\gamma$ is set to 10 for the first 10k steps, then until 100k steps it is linearly reduced to 0.1 and remains fixed thereafter.
Total number of steps is 1000k and it took 10 days on a TESLA V100 GPU card for each model. 

\vspace{-1.5mm}
\subsection{Objective Evaluation in Japanese}
\vspace{-1.5mm}

We first computed F0 distortion between input speech and reconstructed speech as an objective evaluation. We also ran the P.563 algorithm~\cite{rec2004p} for evaluating the reconstruction quality. Results for each test speaker and the average are shown in Table~\ref{tab:rmse-f0}. We can see that the proposed extension reduced the F0 distortion for all test speakers as expected. Moreover, we can see that reconstructed speech using the extended VQ-VAE generally has higher P563 scores than that using the original VQ-VAE. The averaged score of the extended VQ-VAE is 4.2 and is the same as that of the input natural speech. 
%

\begin{table}[t]
\centering
\caption{F0 RMSE errors and P563-based estimated MOS scores of each Japanese test speaker and their average. Natural speech MOS score is 4.2 on average. Original VQ-VAE doesn't have F0 encoder but the extended one does.} 
\label{tab:rmse-f0}
\vspace{-1.5mm}
\footnotesize
\begin{tabular}{c|c|c|c|c}
\hline
%
\multirow{2}{*}{Speaker}  & \multirow{2}{*}{Gender }             & \multirow{2}{*}{VQ-VAE}    & $\log$ F0  & P563  \\
                         &                                       &                            &  RMSE      & MOS  \\\hline                                              
\multirow{2}{*}{Speaker 1}    & \multirow{2}{*}{M}  &  Original       &   0.51   & 3.8   \\ 
                              &                      & Extended &    0.30   & 4.2   \\ \hline
\multirow{2}{*}{Speaker 2}    & \multirow{2}{*}{M}                     & Original       &   0.46   & 3.6   \\ 
                               &                     & Extended &    0.20   & 5.0   \\ \hline    
\multirow{2}{*}{Speaker 3}    & \multirow{2}{*}{M}                    & Original       &   0.49  & 3.8   \\ 
                              &                      & Extended &    0.27   & 3.9   \\ \hline
\multirow{2}{*}{Speaker 4}    &\multirow{2}{*}{M}                      & Original       &   0.43  & 3.7   \\ 
                              &                      & Extended &    0.24   & 3.9   \\ \hline
\multirow{2}{*}{Speaker 5}    &\multirow{2}{*}{F}   & Original       &   0.36   & 4.0   \\ 
                              &                      & Extended &    0.25   & 4.5   \\ \hline
\multirow{2}{*}{Speaker 6}    &\multirow{2}{*}{F}                     & Original       &   0.30   & 3.7   \\ 
                              &                      & Extended &    0.23   & 4.2   \\ \hline
\multirow{2}{*}{Speaker 7}    &\multirow{2}{*}{F}                     & Original       &   0.36   & 3.0   \\ 
                               &                     & Extended &    0.27 & 3.5   \\ \hline

\multirow{2}{*}{Speaker 8}  & \multirow{2}{*}{F}                    &Original       &   0.41   & 4.1   \\ 
                            &                        & Extended &    0.27   & 4.1   \\ \hline
\multirow{2}{*}{Average}   & \multirow{2}{*}{M+F}                   &Original       &   0.42   & 3.8   \\ 
                           &                                         & Extended &    0.26   & 4.2    \\ \hline 
\end{tabular}
\vspace{-6mm}

\end{table}


\vspace{-1.5mm}
\subsection{Subjective Evaluation in Japanese}
\vspace{-1.5mm}

\label{sec:res}

We conducted an AB preference test and compared the VQ-VAE with and without the F0 encoder subjectively\footnote{Audio samples are available at: ~\url{https://nii-yamagishilab.github.io/yi-demo/interspeech-2020/index.html}.}. Subjects are native speakers of Japanese and the number of subjects was 21. Each subject was asked to listen to a total of 20 pairs, and for each pair they were asked to choose the one that sounds better in terms of appropriateness of pitch accents and quality. 

Table~\ref{tab:abtest-japanese} shows the preference score and 95\% confidence intervals for each test speaker, and the average. As we can see from the table, our proposed extension has achieved significantly higher preference scores than the original VQ-VAE for all the test speakers, and this clearly demonstrates that the extended VQ-VAE learned complementary supra-segemental features, which are critical for human perception. 


\begin{table}[t]
\centering
\caption{Preference scores and 95\% confidence intervals (CI) of each Japanese test speaker and their average.}
\label{tab:abtest-japanese}
\vspace{-1.5mm}
\footnotesize
\begin{tabular}{c|c|c|c}
\hline
Speaker    & Original [\%] & Extended [\%] & 95\% CI \\ \hline
Speaker 1  &   8.8      &    91.2      &    $\pm$3.6     \\ 
Speaker 2  &   15.4      &   84.6       &   $\pm$6.7      \\ 
Speaker 3  &    2.4    &    97.6      &   $\pm$3.4      \\ 
Speaker 4  &   6.0      &    94.0      &   $\pm$6.8     \\ 
Speaker 5  &   4.1      &    95.9      &    $\pm$5.7     \\ 
Speaker 6  &   11.3      &     88.7     &   $\pm$8.8      \\ 
Speaker 7  &    7.6     &     92.4     &   $\pm$7.4     \\ 
Speaker 8  &     5.8    &    94.2      &  $\pm$6.6       \\ \hline
Average    &    7.7 &  92.3    & $\pm$2.1 \\ \hline
\end{tabular}
\vspace{-1.5mm}
\end{table}

\vspace{-1.5mm}
\subsection{Extensions to Tonal Languages: Mandarin}

\vspace{-1.5mm}

We also trained a speaker-dependent VQ-VAE without and with the F0 encoder using the Mandarin database to analyze what happens in another language. Table \ref{tab:rmse-f0-mandarin} gives the F0 distortion between input speech and reconstructed speech and P.563-based estimated MOS scores. From this table, we can see that the proposed extension effectively reduced the F0 distortion for the Mandarin speaker, as expected. The RMSE value for the Mandarin speaker is smaller than those of the Japanese speakers, but this is not surprising since the Mandarin model is speaker-dependent whereas the Japanese model is speaker-independent, and all test speakers in Japanese are unseen. 

P.563-based MOS of the Mandarin speaker shows a different tendency from the Japanese results in Table \ref{tab:rmse-f0}. The MOS scores of Mandarin reconstructed speech are 4.4 for the VQ-VAE without and with the F0 encoder, which is the same as that of the input speech. This is because Mandarin syllables and a tone index can be coupled and be represented together, unlike the Japanese pitch accents. For learning tones separately from syllables, explicit disentanglement using adversarial loss or mutual information \cite{hjelm2018learning} would be needed. This is our next step. 

\begin{table}[t]
\centering
\caption{F0 RMSE errors and P563-based estimated MOS scores for Mandarin speaker. Natural speech MOS score is 4.4.}
\label{tab:rmse-f0-mandarin}
\vspace{-1.5mm}
\footnotesize
\begin{tabular}{c|c|c|c}
\hline
\multicolumn{1}{l|}{Speaker} & VQ-VAE       & $\log$ F0 RMSE  & P563 based MOS \\ \hline
\multirow{2}{*}{Chinese}       & Original      &   0.22          &  4.4    \\ 
                               & Extended &   0.15          &  4.4   \\ \hline
\end{tabular}
\vspace{-6mm}
\end{table}

\vspace{-1.5mm}
\section{Conclusions} 
\vspace{-1.5mm}

This paper proposed an extension of VQ-VAE for learning F0-related suprasegmental information additionally. The extended framework uses a F0 trajectory as additional input and learns discrete representations for F0. We constructed a speaker-independent VQ-VAE model using the Japanese databases and showed that reconstructed speech has lower F0 RMSE for all unseen speakers. A listening test also indicated significantly higher preference scores. An additional experiment using the single-speaker Mandarin database also showed the proposed model reduced F0 distortion in another language. Our future work will include the evaluation of down-stream tasks using learned discrete representations such as TTS and explicit disentanglement of latents for F0 and waveforms. 

\vspace{1mm}
\noindent
\textbf{Acknowledgments}
\footnotesize{Cheng-I is supported by the Merrill Lynch Fellowship, MIT. This work was partially supported by a JST CREST Grant (JPMJCR18A6, VoicePersonae project), Japan, and by MEXT KAKENHI Grants (16H06302, 18H04112, 18KT0051, 19K24373 Japan. The numerical calculations were carried out on the TSUBAME 3.0 supercomputer at the Tokyo Institute of Technology.} 

\newpage 

\bibliographystyle{IEEEtran}
\bibliography{mybib}

\end{document}